\documentclass[10pt,a4paper]{article}
\usepackage[utf8]{inputenc}
\usepackage{amsmath}
\usepackage{amsfonts}
\usepackage{amssymb}
\usepackage{graphicx}
\usepackage{csquotes}
\MakeOuterQuote{"}
\author{Carringtone Kinyanjui, Dismas S. Wamalwa}
\title{SCHWARZSCHILD SOLUTION OF THE MODIFIED EINSTEIN FIELD EQUATIONS}
\begin{document}
\maketitle
\begin{center}
Department of Physics\\
University of Nairobi\\
Chiromo, Nairobi, 30197-0100, KENYA
\end{center}
\begin{abstract}
A reformulation  of the Schwarzschild solution of the linearized Einstein field equations in post-Riemannian Finsler spacetime is derived. The solution is constructed in three stages: the exterior solution, the event-horizon solution and the interior solution. It is shown that the exterior solution is asymptotically similar to Newtonian gravity at large distances implying that Newtonian gravity is a low energy approximation of the solution. Application of Eddington-Finklestein coordinates is shown to reproduce the results obtained from standard general relativity at the event horizon. Further application of Kruskal-Szekeres coordinates reveals that the interior solution contains maximally extensible geodesics.

\textbf{AMS Subject Classification:} 83C05, 83C57, 83C75

\textbf{Key Words and Phrases:} Einstein field equations, Schwarzschild metric, Black holes, Finsler spaces
\end{abstract}
\section{Introduction}
The general theory of relativity as proposed by Einstein \cite{Einstein} relates the curvature of spacetime to gravity. 
The theory describes the relation between the curvature of spacetime to the energy of an object. This can  succinctly be described by Einstein field equations (stated without the cosmological constant) \cite{Mahon}:
\begin{equation*}
R_{\mu \nu}-\frac{1}{2}g_{\mu \nu}R=\frac{8\pi G}{c^4}T_{\mu \nu}
\end{equation*}
where $R_{\mu \nu}$ is the Ricci curvature tensor, $R$ is the Ricci scalar, $g_{\mu \nu}$ is the metric, G is the gravitational constant, $T_{\mu \nu}$ is the stress energy tensor and $c$ is the speed of light. The theory was soon tested observationally by Eddington in 1919 and found to be correct \cite{Dyson}.
The simplest analytical solution to the field equations is the solution for a static uncharged and spherical mass. The solution was proposed by Karl Schwarzschild in 1916 \cite{Schwarzschild}.
The Schwarzschild metric for paths along radial lines is given by \cite{Mahon}:
\begin{equation}
d\mathit{s}^2=\left(1-\frac{2m}{\mathit{r}}\right)^{-1}d\mathit{t}^2+\left(1-\frac{2m}{\mathit{r}}\right)d\mathit{r}^2
\end{equation} 
It can be seen that the metric is singular at $\mathit{r}=2m$ and $\mathit{r}=0$ \cite{Hobson}. However, a change of coordinates particularly proposed by Eddington \cite{Eddington} and later by Finklestein \cite{Finkelstein}, showed that the singularity at $\mathit{r}=2m$ can be  removed. However, the curvature singularity due to spacetime structure at $\mathit{r}=0$ persists and can not be removed \cite{Hobson}.

The interior solution ($\mathit{r}<2m$) describes objects called "blackholes"  that "swallow" objects that come too close to them ($\mathit{r}=2m$) and have escape velocities greater than the speed of light. This means that at $\mathit{r}<2m$, there is no possibility of escape \cite{Mahon}. While scientific speculation on the existence of black holes predates general relativity \cite{Montgomery}, they seem implicit within the description of the theory. Initially, blackholes were not accepted as physically feasible objects. According to Einstein, the solution was in fact only a mathematical curiosity and of no astrophysical importance \cite{Bartusiak}. Chandrasekhar \cite{Chandrasekhar}, showed that beyond some mass limit, stars at the endpoints of stellar evolution and therefore undergoing gravitational collapse can not be held back by electron degeneracy. Later work by Oppenheimer, Volkoff and Tolmann  generalized Chandrasekhar's work \cite{Oppenheimer} and proved that collapse to a blackhole is astrophysically feasible. Since then, the reality of astrophysical black holes has been confirmed \cite{Webster}, \cite{Johnston}.

Important theoretical work in  the description of blackholes, both from general relativity and from a quantum field theoretic perspective has been done by Hawking et. al. \cite{Ford}, \cite{Senovilla}. It is hoped that quantum gravity will resolve the singularity problem in the centre of stationary blackholes. Gambini and Pullin \cite{Gambini}, \cite{Mahon 2009} have proposed that Loop Quantum Gravity can lead to the description of a non-singular quantised Schwarzschild metric .
In this work, we shall take a different approach from the (relatively) standard gravity quantisation procedure as a correction of gravity at the scale of black hole energies. We shall rely on the results of the model formulated within the frame work of finsler spaces. Finsler spaces \cite{Shen}, \cite{Asanov} can be thought of as generalizations of Riemann spaces. In Particular, we adopt the extended Einstein field equations in Finsler geometry \cite{Wamalwa}: 
\begin{equation}
\left( R_{\mu\nu}-\frac{1}{2}g_{\mu \nu}R\right)+\left(S_{\mu\nu}-\frac{1}{2}g_{\mu \nu}S\right)=\frac{8\pi G}{c^4}\left(T_{\mu \nu}+\tau_{\mu\nu}\right)
\end{equation} 
where $S_{\mu\nu}$ and $S$ are additional Ricci tensor and Ricci scalar terms respectively. Furthermore, $S_{\mu\nu}$ and $R_{\mu\nu}$ are functions of position, $\mathit{x}$ and velocity, $\mathit{y}$ respectively. We hold the velocity terms constant so that $R_{\mu\nu}(\mathit{x,y})=R_{\mu\nu}(\mathit{x})$ and $S_{\mu\nu}(\mathit{x,y})=S_{\mu\nu}(\mathit{x})$.
It should be noted that the field equations in equation (2) above are difficult to solve. We shall therefore proceed to solve the above field equations with an extra constraint of symmetry and hence develop a mathematical model of Schwarzschild blackholes appropriately. In this respect, we shall assume that components of the tensors  $R_{\mu\nu}$ and $S_{\mu \nu}$ are related through a symmetrical linear transformation to be described later.

\section{The Vacuum Field Equations}
We can rewrite Einstein field equations in equation (2) above as:
\begin{equation}
Z^{\alpha \beta } -\frac{1}{2}Zg^{\alpha\beta}=kQ^{\alpha\beta}
\end{equation}
where 
\begin{equation}
Z^{\alpha \beta}=R^{\alpha \beta}+S^{\alpha \beta}
\end{equation}
\begin{equation}
Z=R+S
\end{equation}
\begin{equation}
Q^{\alpha\beta}=T^{\alpha \beta}+\tau^{\alpha\beta}
\end{equation}
Transforming equation (3) into covariant form, we obtain
%
%
\begin{equation}
Z_{\mu \nu}-\frac{1}{2}Z g_{\mu \nu}=kQ_{\mu \nu} 
\end{equation}
We shall now proceed to find the vacuum field equations corresponding to equation (7).
Setting $Q_{\mu \nu}=0$  and performing contraction with metric tensor, it can be shown that
\begin{equation}
R_{\mu \nu}+S_{\mu \nu}=0
\end{equation}
These are our desired vacuum field equations. We next consider a solution of the vacuum field equations for
the Schwarzschild metric.
\subsection{Solution of the vacuum field equations}
The difficulty of solving Finsler extended Einstein field equations is evident. In this paper, we simply introduce a further constraint of symmetry by demanding that the tensor terms $R_{\mu \nu}$ and $S_{\mu \nu}$ are linearly related. While this of course limits the richness of the theory of Finsler spaces, it helps us develop a physically relevant and realistic model.
In order to solve equation (8), we follow the standard procedure of finding the actual form of the metric by constraining its functional parameters. 
Rewriting equation (8) for $\mu=\nu=\theta$, we have:
\begin{align}
R_{\hat{\theta}\hat{\theta}}+S_{\hat{\theta}\hat{\theta}}=0
\end{align}
It has been shown that the value of $R_{\hat{\theta}\hat{\theta}}$ [2] is:
\begin{align}
R_{\hat{\theta}\hat{\theta}}=\frac{2}{\mathit{r}}\frac{d\lambda}{d\mathit{r}}e^{-2\lambda}+\frac{1-e^{-2\lambda}}{\mathit{r}^2}
\end{align}
But $R_{\hat{\theta}\hat{\theta}}$ and $S_{\hat{\theta}\hat{\theta}}$ are linearly related, and therefore similar in structure. Indeed given two differential equations $D_1(x)$ and $D_2(y)$  such that:
\begin{align*}
 D_1(\mathit{x})+D_2(\mathit{y})=0;\quad \mathit{y}=\alpha \mathit{x}\\
\mbox{or}\;\; D_1(\mathit{x})+D_2(\alpha \mathit{x})=0\\
\mbox{therefore,}\;\; D_1(\mathit{x})=-\alpha D_2(\mathit{x})\\
\end{align*}
implying that:
\begin{equation}
\frac{D_1(\mathit{x})}{D_2(\mathit{x})}=-\alpha=\gamma
\end{equation}
It thus, follows that $D_1(\mathit{x})$, $D_2(\mathit{x})$ are linearly related and similar in structure. Differential equations can be modelled as matrix eigenvalue problems whereby the differential operator becomes the  matrix and the solution the eigenfunction. Taking into account that for any eigenvalue problem
\begin{equation*}
\hat{A}\textbf{X}=\beta \textbf{X},
\end{equation*}
and vector $\textbf{Y}=g\textbf{X}$, we have 
\begin{equation*}
\hat{A}\textbf{Y}=\beta\textbf{Y} 
\end{equation*}
where g is a constant.
Implication of equation (11) is that the structure of $S_{\hat{\theta}\hat{\theta}}$ arises as:
\begin{equation}
S_{\hat{\theta}\hat{\theta}}=\frac{2}{\mathit{r}}\frac{d\mathit{L}}{dr}e^{-2\mathit{L}}+\frac{1-e^{-2\mathit{L}}}{\mathit{r}^2}
\end{equation}
where $\mathit{L}$ is a function in space that is linearly related to $\lambda$. 
Equation (9) is, therefore, rewritten as:
\begin{equation}
\frac{2}{\mathit{r}}\frac{d\lambda}{d\mathit{r}}e^{-2\lambda}+\frac{1-e^{-2\lambda}}{\mathit{r}^2}+\frac{2}{\mathit{r}}\frac{d\mathit{L}}{d\mathit{r}}e^{-2\mathit{L}}+\frac{1-e^{-2\mathit{L}}}{\mathit{r}^2}=0
\end{equation}
Or
\begin{equation}
2\mathit{r}\frac{d\lambda}{d\mathit{r}}e^{-2\lambda}+1-e^{-2\lambda}+2\mathit{r}\frac{d\mathit{L}}{d\mathit{r}}e^{-2\mathit{L}}+1-e^{-2\mathit{L}}=0
\end{equation}
Application of chain rule
\begin{equation}
\frac{d\mathit{L}}{d\lambda}=\frac{d\mathit{L}}{d\lambda}\frac{d\lambda}{d\mathit{r}}
\end{equation}
and trying to find $\frac{dL}{d\lambda}$ by assuming the most general case of  "pseudopolynomial" functions, we have:
\begin{equation*}
\mathit{L(r)}=A\mathit{r}^n+B\mathit{r}^{n-1}+C\mathit{r}^{n-2}+\cdots+D\mathit{r}+E+F\mathit{r}^{-1}+G\mathit{r}^{-2}+H\mathit{r}^{-3}+\cdots+I\mathit{r}^{-n}
\end{equation*}
\begin{equation*}
\lambda(\mathit{r})=J\mathit{r}^n+K\mathit{r}^{n-1}+L\mathit{r}^{n-2}+\cdots+M\mathit{r}+N+O\mathit{r}^{-1}+P\mathit{r}^{-2}+Q\mathit{r}^{-3}+\cdots+R\mathit{r}^{-n}
\end{equation*}
where $A$ to $R$ are distinct constants.
Therefore,
\begin{equation*}
\frac{d\mathit{L}}{d\lambda}=\frac{A'\mathit{r}^{n-1}+B'\mathit{r}^{n-2}+C'\mathit{r}^{n-3}+\cdots+D+F'\mathit{r}^{-2}+G'\mathit{r}^{-3}+H'\mathit{r}^{-4}+\cdots I'\mathit{r}^{-n-1}}{J'\mathit{r}^{n-1}+K'\mathit{r}^{n-2}+L'\mathit{r}^{n-3}+\cdots+M+O'\mathit{r}^{-2}+P'\mathit{r}^{-3}+Q'\mathit{r}^{-4}+\cdots+R'\mathit{r}^{-n-1}}
\end{equation*}
where primed constants are new constants obtained after differentiation.
Neglecting higher order terms for large $\mathit{r}$, we get:
\begin{equation*}
\frac{d\mathit{L}}{d\lambda}=\frac{A'\mathit{r}^{n-1}+B'\mathit{r}^{n-2}+C'\mathit{r}^{n-3}+\cdots+D+F'\mathit{r}^{-1}}{J'\mathit{r}^{n-1}+K'\mathit{r}^{n-2}+L'\mathit{r}^{n-3}+\cdots+M}
\end{equation*}
We can express $\frac{d\mathit{L}}{d\lambda}$  as:
\begin{equation}
\frac{d\mathit{L}}{d\lambda}=\frac{b(\mathit{r})}{\mathit{r}}
\end{equation}
where 
\begin{equation}
b(\mathit{r})=\frac{A'\mathit{r}^{n}+B'\mathit{r}^{n-1}+C'\mathit{r}^{n-2}+\cdots+D\mathit{r}+E'\mathit{r}^{-1}}{J'\mathit{r}^{n-1}+K'\mathit{r}^{n-2}+L'\mathit{r}^{n-3}+\cdots+M}
\end{equation}
Equations (16) when used in equation (13) yields:
\begin{equation}
2-e^{-2\lambda}+2\mathit{r}\frac{d\lambda}{d\mathit{r}}e^{-2\lambda}+-e^{-2\mathit{L}}+2\mathit{r}\frac{b}{\mathit{r}}\frac{d\lambda}{d\mathit{r}}e^{-2\mathit{L}}=0
\end{equation}
Since $\mathit{L}=k\lambda$, we have equation (18) as:
\begin{equation}
2-e^{-2\lambda}+2\mathit{r}\frac{d\lambda}{d\mathit{r}}e^{-2\lambda}+-e^{-2k\lambda}+2\mathit{r}\frac{b}{\mathit{r}}\frac{d\lambda}{d\mathit{r}}e^{-k2\lambda}=0
\end{equation}
or expressing $k$ as the logarithm of a certain constant $a$, we may write
\begin{equation}
e^{-k2\lambda}=\left( e^{-2\lambda}\right)^k=ae^{-2\lambda}
\end{equation}
Using equation (20) in equation (19) and regarding the small constant $a$ as unity, we have: 
\begin{equation}
-\frac{d\lambda}{d\mathit{r}}2e^{-2\lambda}(\mathit{r}+b)+2e^{-2\lambda}=2
\end{equation}
Equation (21) can easily be integrated to give:
\begin{equation}
e^{-2\lambda}= \left( 1-\frac{2m}{\mathit{r}+b}\right)
\end{equation}
Where $m=\frac{GM}{c^2}$
\section{Modified Schwarzschild Metric}
The standard Schwarzschild metric is written as:
\begin{equation}
d\mathit{s}^2=e^{-2\lambda}d\mathit{t}^2-e^{2\lambda}d\mathit{r}^2-r^2(d\theta^2+sin^2\phi^2)
\end{equation}
If we consider paths along radial lines for light cones around the singularity $\mathit{r}=2m$, the above metric reduces to:
\begin{equation}
d\mathit{s}^2=e^{-2\lambda}d\mathit{t}^2-e^{2\lambda}d\mathit{r}^2
\end{equation}
Using equation (22) in equation (24) yields:
\begin{equation}
d\mathit{s}^2=\left( 1-\frac{2m}{\mathit{r}+b}\right)d\mathit{t}^2-\left( 1-\frac{2m}{\mathit{r}+b}\right)^{-1}d\mathit{r}^2
\end{equation}
Comparison with equation (1) shows that as a result of our computation, there is a correction to the term $2m/\mathit{r}$ present in equation (25).

Let us now consider the asymptotic behaviour of equation (25) at the Schwarzschild radius, at the centre of the black hole and at large distances.
\subsection{Asymptotic behaviour of the metric}
\subsubsection{Behaviour of the metric at r=2m}
At $\mathit{r}=2m$, a radius is defined such that nothing that goes into the blackhole ever gets out \cite{Schwarzschild}. In relativistic terms, the light cones of test particles are completely tipped over such that the geodesics are pointing towards the center of the black hole \cite{Eddington} i.e.,
\begin{equation}
d\mathit{s}^2=0
\end{equation} 
Singularity at $\mathit{r}=2m$ is due to poor choice of coordinate system hence, is called coordinate singularity.
Transformation into Eddington-Finklestein coordinates removes the singularity. We now proceed to transform the metric into Eddington-Finklestein coordinates. These coordinates describe spacetime at the event horizon. They are derived from null geodesics where the metric is set to zero (equation 26).
Combining equations (26) and (25), we obtain:
\begin{equation*}
\left( 1-\frac{2m}{\mathit{r}+b}\right)d\mathit{t}^2=\left(1-\frac{2m}{\mathit{r}+b}\right)^{-1}d\mathit{r}^2
\end{equation*}
This equation can be solved to give:
\begin{eqnarray*}
t=\mathit{r}+2m\;ln(\frac{\mathit{r}+b}{2m}-1)+c\\
\end{eqnarray*}
where $c=b-2m+2m\;ln\;2m$ is a constant.
Relabelling $\mathit{t}$ as $\mathit{r*}$, we have:
\begin{equation}
\mathit{r*}=\mathit{r}+2m\;ln(\frac{\mathit{r}+b}{2m}-1)+c
\end{equation}
so that
\begin{equation}
\frac{d\mathit{r}*}{d\mathit{r}}=\frac{\mathit{r}+b}{\mathit{r}+b-2m}
\end{equation}
Introducing "tortoise" coordinates coordinates [2] of the form:
\begin{eqnarray}
\mathit{v}=\mathit{t}+\mathit{r*}\\
\mathit{u}=\mathit{t}-\mathit{r*},
\end{eqnarray}
it can easily be shown that:
\begin{equation}
d\mathit{r}^2=\frac{1}{4}\left(\frac{\mathit{r}+b-2m}{\mathit{r}+b}\right)^2(d\mathit{v}^2-2d\mathit{u}d\mathit{v}+d\mathit{v}^2)\\
\end{equation}
and
\begin{equation}
d\mathit{t}^2=\frac{1}{4}(d\mathit{u}^2+2d\mathit{u}d\mathit{v}+d\mathit{v}^2)
\end{equation}
Using equations (22), (31) and (32) in equation (25), we obtain:
\begin{equation*}
d\mathit{s}^2=\left(1-\frac{2m}{\mathit{r}+b}\right)\cdot \frac{1}{4}(d\mathit{u}^2+2d\mathit{u}d\mathit{v}+d\mathit{v}^2)-
\end{equation*}
\begin{equation}
\left(1-\frac{2m}{\mathit{r}+b}\right)^{-1}\frac{1}{4}\left(1-\frac{2m}{\mathit{r}+b}\right)^2(d\mathit{v}^2-2d\mathit{u}d\mathit{v}+d\mathit{v}^2)
\end{equation}
Equation (33) can be reduced to:
\begin{eqnarray}
d\mathit{s}^2=\left(1-\frac{2m}{\mathit{r}+b}\right)dudv
\end{eqnarray}
Equation (34) is the metric restated in Eddington-Finklestein coordinates. Taking asymptotic behaviour at $\mathit{r}=2m$ and noting that $2m>>b$ yields:
\begin{equation}
d\mathit{s}^2|_{\mathit{r}=2m}=0
\end{equation}
which is consistent with equation (25). The result is also consistent with the behaviour of the standard Schwarzschild metric \cite{Mahon} and hence, with the mathematical predictions of standard general relativity. Therefore, the theory is asymptotically similar to the standard theory outside the black hole and at the event horizon.
\subsubsection{Asymptotic behaviour inside the Scwhwarzschild black hole }
By inspection, it can  easily be seen that the metric in equation (25) is non-singular at $\mathit{r}=0$ i.e.,
\begin{equation}
d\mathit{s}^2|_{\mathit{r}=0}=(1-\frac{2m}{b})d\mathit{t}^2+(1-\frac{2m}{b})^{-1}d\mathit{r}^2
\end{equation}
However, to be completely sure that we have eliminated the singularity, we need to transform the metric into Kruskal-Szekeres coordinates \cite{Hobson}. Using equations (29) and (30) in equation (27), we get: 
\begin{equation}
\mathit{r*}=\mathit{r}+2m\;ln(\frac{\mathit{r}+b}{2m}-1)+c=\frac{1}{2}(\mathit{v}-\mathit{u})
\end{equation}
Dividing equation (37) by $2m$, we obtain:
\begin{equation}
r+b-2m=2m\;e^{{\frac{1}{4m}(v-u)}{ -\frac{1}{2m}(r+c)}}
\end{equation}
Rearranging equation (34) and applying equation (38), we obtain:
\begin{equation*}
d\mathit{s}^2=\frac{2m}{\mathit{r}+b}e^{\frac{(\mathit{v}-\mathit{u})}{4m}}e^{-\frac{(\mathit{r}+c)}{2m}}d\mathit{u}d\mathit{v}
\end{equation*}
Using transformations of the form:
\begin{equation*}
\mathit{U}=-e^{-\frac{\mathit{u}}{4m}};\qquad \mathit{V}=e^{-\frac{\mathit{v}}{4m}},
\end{equation*}
it is easy to show that:
\begin{eqnarray*}
d\mathit{s}^2=\frac{32m^3}{\mathit{r}+b}e^{-\frac{(\mathit{r}+c)}{2m}}d\mathit{U}d\mathit{V}
\end{eqnarray*}
so that application of cooordinate transformation suggested by Kruskal \cite{Schwarzschild} gives:
\begin{equation}
d\mathit{s}^2=\frac{32m^3}{\mathit{r}+b}e^{-\frac{(\mathit{r}+c)}{2m}}(d\mathit{T}^2-d\mathit{X}^2)
\end{equation}
Taking asymptotic behaviour at $\mathit{r}=0$, we obtain
\begin{equation}
d\mathit{s}^2|_{\mathit{r}=0}=\frac{32m^3}{b}e^{-\frac{\mathit{r}+c}{2m}}(d\mathit{T}^2-d\mathit{X}^2)
\end{equation}
showing that the metric has non-singular behaviour at $\mathit{r}=0$.
\subsubsection{Asymptotic behaviour at large distances}
At large distances, equation (25) reduces to:
\begin{equation}
d\mathit{s}^2=d\mathit{t}^2-d\mathit{r}^2
\end{equation}
Equation (41) is just the Minkowski spacetime \cite{Hobson}. At large distances from the mass, curvature is minimised, the general relativistic curvature corrections are absent and thus the metric is asymptotically flat and similar to the standard metric.

\subsection{Conclusion}
The metric corresponding to a Schwarzschild solution for the extended Einstein field equations has been derived. The metric has been shown to have the external Schwarzschild solution as an aymptotic extension at long distances. At the event horizon, the metric is shown to be equal to the standard Schwarzschild metric. Further, and more interestingly, the metric is shown to be non-singular at $\mathit{r}=0$. We invite further exploration of the work presented, including the calculation of the Kretschmann invariant and possible modification of the Kerr metric. We hope in later work to explore a unification scheme based on the extended field equations that will assist in the determination of the constant b.



\begin{thebibliography}{99}
\bibitem{Einstein}  A. Einstein,The Field Equations of Gravitation, {\it Proceedings of The
Prussian Academy of Sciences}, {\bf 2}, (1915), 884-887

\bibitem{Mahon} D. McMmahon, {\it Relativity Demystified}, McGraw-Hill, USA(2006). 

\bibitem{Dyson} F.W. Dyson, A.S. Eddington, C.R. Davidson, A Determination of the Deflection
of Light by the Sun’s Gravitational Field, from Observations Made at the Total
eclipse of May 29, 1919, {\it Philosophical Transactions of The Royal Society of London}
{\bf 220}, (1919), 291-333, doi:10.1098/rsta.1920.0009.

\bibitem{Schwarzschild} K. Schwarzschild, On the Gravitational Field of a Mass Point by Einstein’s
Theory, {\it Proceedings of The Prussian Academy of Sciences}, (1916), 189-196 , arXiv:physics/9905030 [physics.hist-ph]

\bibitem{Hobson} M. P. Hobson , G. P. Efstathiou, A. N. Lasenby, {\it General Relativity: An Intro
duction for Physicists}, Cambridge University Press, UK(2006), 196-220.\\
 
\bibitem{Eddington} A. Eddington, A Comparison of Whitehead’s and Einstein’s
Formulae, {\it Nature\\(Letters to The Editor)}, {\bf 113}, 2832(1924), 192.

\bibitem{Finkelstein} D. Finkelstein, Past-Future Asymmetry of the Gravitational Field of a Point
Particle, {\it Physical Review Journal}, {\bf 110}, N0.4(1958), doi:10.1103/PhysRev.110.965.\\

\bibitem{Montgomery} C.Montgomery, W.Orchiston, I. Whittingham, Michell, Laplace and the Origin of
the Black Hole Concept, {\it Journal of Astronomical History and Heritage}, {\bf 12}, 2(2009), 90-96\\

\bibitem{Bartusiak} M. Bartusiak, {\it Blackholes: How an Idea abandoned by Newtonians, hated by
Einstein, and gambled on by Hawking, became loved}, Yale
University Press, USA(2016)\\ 

\bibitem{Chandrasekhar} S. Chandrasekhar, The Maximum Mass of Ideal White Dwarfs, {\it Astrophysical
Journal}, {\bf 74}, (1931), 81-82

\bibitem{Oppenheimer} J.R. Oppenheimer, G.M. Volkoff, On Massive Neutron Cores, {\it Physical
Review}, {\bf 55}, 4(1939), 374–381, doi:10.1103/PhysRev.55.374

\bibitem{Webster} B.L. Webster,P. Murdin, Cygnus X-1—a Spectroscopic Binary with a Heavy
Companion?, {it Nature(Letters to Nature)}, {\bf 235}, 37-38(1972), 235. 

\bibitem{Johnston} R. Johnston, List of black hole candidates, {\it http://www.johnstonsarchive.net/
\\relativity/bhctable.html}, (2016), Accessed on 12th July 2016\\

\bibitem{Ford} L.H. Ford, Quantum Field Theory in Curved Spacetime, {\it arXiv:gr-qc/9707062}, (1997).

\bibitem{Senovilla} J.M.M. Senovilla, Singularity Theorems in General Relativity: Achievements
and Open Questions, {\it http://arxiv.org/abs/physics/0605007v1}, (2006).

\bibitem{Gambini} R. Gambini, J. Pullin, Loop Quantisation of The Schwarzschild
Blackhole,{\it Physical Review Letters}, {\bf 110}, 211301(2013), doi:10.1103/PhysRevLett.110.211301

\bibitem{Mahon 2009} D. McMmahon, {\it String Theory Demystified}, McGraw-Hill, New York(2009),
239-255

\bibitem{Shen} Z. Shen, {\it Lectures on Finsler Geometry}, Indiana University-Purdue
University Indianapolis, USA(2001)

\bibitem{Asanov} G.S. Asanov, {\it Finsler Geometry, Relativity and Gauge Theories},
  D Reidei Publishing Company, Holland(1985)

\bibitem{Wamalwa} D.S.Wamalwa, J.A. Omolo, Generalised Relativistic Dynamics In A Non-Inertial
Frame of Reference, {\it Indian Journal of Physics}, {\bf 84}, N0.9(2010), 1241-1255

\end{thebibliography}
\end{document}